\documentclass[12pt,oneside]{article}

\usepackage[left=3.4cm,top=3.6cm,bottom=3.6cm,right=3.4cm,head=0cm,foot=0.7cm]{geometry}

\usepackage{amsmath,amssymb}

\newcommand{\field}[1]{\mathbb{#1}}

\date{}

\title{ {\bf Boltzmannian Equilibrium in \\ Stochastic Systems}  \vspace{5mm} }

\author{ {\bf Charlotte Werndl}\\ Department of Philosophy, University of Salzburg\\Department of Philosophy, Logic and Scienific Method, LSE\\ charlotte.werndl@sbg.ac.at\\  \\ {\bf Roman Frigg} \\Department of Philosophy, Logic and Scientific Method, LSE\\ Centre for Philosophy of Natural and Social Science, LSE\\ r.p.frigg@lse.ac.uk\\ \\ {\small forthcoming in:} {\small Michela Massimi and Jan-Willem Romeijn (eds):} \\  {\small Proceedings of the EPSA15 Conference. Springer.}}

\begin{document}

\newtheorem{theorem}{Theorem}

\maketitle

\author

\begin{abstract}
\noindent Equilibrium is a central concept of statistical mechanics. In previous work we introduced the notions of a Boltzmannian $\alpha$-$\varepsilon$-equilibrium and a Boltzmannian $\gamma$-$\varepsilon$-equilibrium (Werndl and Frigg 2015a, 2015b). This was done in a deterministic context. We now consider systems with a stochastic micro-dynamics and transfer these notions from the deterministic to the stochastic context. We then prove stochastic equivalents of the Dominance Theorem and the Prevalence Theorem. This establishes that also in stochastic systems equilibrium macro-regions are large in requisite sense.
\end{abstract}


\section{Introduction}

Equilibrium is a central concept of statistical mechanics. In Boltzmannian statistical mechanics (BSM) equilibrium is standardly associated with the largest macro-region, where macro-regions are parts of the accessible phase space consisting of micro-states that are the supervenience base for the same macro-state. In two recent papers we argue that the standard picture lacks a foundation and ought to be replaced by an alternative approach (Werndl and Frigg 2015a, 2015b). We develop this approach in detail under the assumption that the underlying micro-dynamics is deterministic. In this paper we give up this assumption and generalise our approach to systems with a stochastic micro-dynamics.\\

In Section \ref{Rethought} we introduce the main pillars of our programme. In Section \ref{Stochastic} we present stochastic systems. In Section \ref{S-Equilibrium} we carry over our key concepts from the deterministic to the stochastic context and formulate the main theorems, which we prove in the Appendix. In Section \ref{Example} we illustrate our claims with the example of the lattice gas, an important and widely used model in physics. In Section \ref{Conclusion} we summarise our results and add some concluding remarks.


\section{Boltzmannian Equilibrium Rethought}\label{Rethought}

In this section we briefly present the new definition of equilibrium we proposed in previous work (Werndl and Frigg 2015a, 2015b). Consider a system consisting of $n$ particles in an isolated and bounded container. The system's micro-state is a point $x$ in its $6n$-dimensional state space $\Gamma$. The system's dynamics is given by a deterministic time evolution  $\phi_{t}$, where $\phi_{t}(x)$ is the state into which $x \in \Gamma$ evolves after $t$ time steps. The system's energy is preserved and so the system's motion is confined to the energy hypersurface $\Gamma_{E}$. The hypersurface is equipped with a sigma algebra $\Sigma_{E}$ and a normalised measure $\mu_{E}$ which is invariant under $\phi_{t}$. Taken together these elements constitute the measure-preserving deterministic dynamical system $(\Gamma_{E},\Sigma_{E},\mu_{E},\phi_{t})$.\\

From a macroscopic point of view the system can be characterised by a set $\{v_{1}, ..., v_{k}\}$ of macro-variables ($k \in \field{N}$). The $ v_{i}$ are functions on $\Gamma_{E}$ that assume values in the range $\field{V}_{i}$, and capital letters $V_{i}$ denote the values of $v_{i}$. A particular set of values $\{V_{1}, ..., V_{k}\}$ defines a \emph{macro-state} $M_{V_{1}, \ldots, V_{k}}$. A set of macro-states is complete iff (if and only if) it contains all states a system can be in. In Boltzmannian statistical  mechanics macro-states supervene on micro-states and hence every macro-state $M$ is associated with a macro-region $\Gamma_{M}$ consisting of all $x\in\Gamma_{E}$ for which the system is in $M$. For a complete set of macro-states the $\Gamma_{M}$ form a partition of $\Gamma_{E}$.\newline

The equilibrium macro-state is $M_{eq}$ and its macro-region is $\Gamma_{M_{eq}}$. A crucial aspect of the standard presentation of BSM is that $\Gamma_{M_{eq}}$ is the largest macro-region. The notion of the `largest macro-region' can be interpreted in two ways. The first takes `largest' to mean that the equilibrium macro-region takes up a large part of $\Gamma_{E}$. We say that $\Gamma_{M_{eq}}$ is \emph{$\beta$-dominant} iff $\mu_{E}(\Gamma_{M_{eq}}) \geq\beta$ for a particular $\beta\in (\frac{1}{2},1]$. If $\Gamma_{M_{eq}}$ is $\beta$-dominant, then it is in fact also $\beta'$-dominant for all $\beta'$ in $(1/2, \, \beta)$. The second reading takes `largest' to mean `larger than any other macro-region'. We say that $\Gamma_{M_{eq}}$ is \emph{$\delta$-prevalent} iff $\min_{M \neq M_{eq}} [\mu_{E}(\Gamma_{M_{eq}}) -\mu_{E}(\Gamma_{M})]\geq\delta$ for a particular real number $\delta > 0$. This implies that if $\Gamma_{M_{eq}}$ is $\delta$-prevalent, then it is also $\delta'$-prevalent for all  $\delta'$ in $(0, \, \delta)$. We do not adjudicate between these different definitions - either meaning of `large' can be used to define equilibrium. It ought to be pointed out, however, that they are not equivalent: whenever an equilibrium macro-region is $\beta$-dominant, there exists a range of values for $\delta$ so that the macro-region is also $\delta$-prevalent for these values, but the converse fails.\\

The question now is: why is the equilibrium state $\beta$-dominant or $\delta$-prevalent? A justification ought to be as close as possible to thermodynamics. In thermodynamics a system is in equilibrium when change has come to a halt and all thermodynamic variables assume constant values (cf. Reiss 1996, 3). This would suggest a definition of equilibrium according to which every initial condition lies on a trajectory for which $\{v_{1}, ..., v_{k}\}$ eventually assume constant values. Yet this is unattainable for two reasons. First,  the values of the $v_{i}$ will never cease to fluctuate due to Poincar\'{e} recurrence. This, however, is no cause for concern. Experimental results show that systems exhibit fluctuations away from equilibrium (Wang et al.\ 2002), and hence the strict thermodynamic notion of equilibrium is actually \emph{unphysical}. Second, in dynamical systems we cannot expect \textit{every} initial condition to approach equilibrium (see, for instance, Callender 2001). For these reasons we define equilibrium as the macro-state in which trajectories starting in most initial conditions spend most of their time.\\

To make this idea precise we introduce the long-run fraction of time a system spends in a region $A\in \Sigma_{E}$ when the system starts in micro-state $x$ at time $t=0$:

\begin{eqnarray}\label{LF}
LF_{A}(x)&=&\lim_{t\rightarrow\infty}\frac{1}{t}\int_{0}^{t}1_{A}(\phi_{\tau}(x))d\tau\,\,\textnormal{for continuous time, i.e.}\,\,t\in\field{R},\,\,\\
LF_{A}(x)&=&\lim_{t\rightarrow\infty}\frac{1}{t}\sum_{\tau=0}^{t-1}1_{A}(\phi_{\tau}(x))\,\,\textnormal{for discrete time, i.e.}\,\,t\in\field{Z},\nonumber
\end{eqnarray}

\noindent where $1_{A}(x)$ is the characteristic function of $A$, i.e.\ $1_{A}(x)=1$ for $x\in A$ and $0$ otherwise.\\

The notion `most of their time' is beset with the same ambiguity as the `largest macro-state'. On the first reading most of the time means more than half of the total time. This leads to the following formal definition of equilibrium:

\begin{quote}

\textit{BSM $\alpha$-$\varepsilon$-Equilibrium.} Consider an isolated system $S$ whose macro-states are specified in terms of the macro-variables $\{v_{1}, ..., v_{k}\}$ and which, at the micro level, is a measure-preserving deterministic dynamical system $(\Gamma_{E},\Sigma_{E},\mu_{E},\phi_{t})$. Let $\alpha$ be a real number in $(0.5, 1]$, and let $1 \gg \varepsilon \ge 0$ be a very small real number. If there is a macro-state $M_{V_{1}^{*}, ..., V_{k}^{*}}$ satisfying the following condition, then it is the $\alpha$-$\varepsilon$-equilibrium state of $S$: There exists a set $Y\subseteq \Gamma_{E}$ such that $\mu_{E}(Y)\geq 1-\varepsilon$, and all initial states $x\in Y$ satisfy
\begin{equation}
LF_{\Gamma_{M_{V_{1}^{*}, ..., V_{l}^{*}}}}\!(x) \, \geq \, \alpha.
\end{equation}
We then write $M_{\alpha\textnormal{-}\varepsilon\textnormal{-}eq}\, := \, M_{V_{1}^{*}, ..., V_{k}^{*}}$.
\end{quote}

\noindent An obvious question concerns the value of $\alpha$. Often the assumption seems to be that $\alpha$ is close to one. This is reasonable but not the only possible choice. For our purposes nothing hangs on a particular choice of $\alpha$ and so we leave it open what the best choice would be.\\

On the second reading `most of the time' means that the system spends more time in the equilibrium macro-state than in any other macro-state. This idea can be rendered precise as follows:

\begin{quote}
\textit{BSM $\gamma$-$\varepsilon$-Equilibrium.}
Consider an isolated system $S$ whose macro-states are specified in terms of the macro-variables $\{v_{1}, ..., v_{k}\}$ and which, at the micro level, is a measure-preserving deterministic dynamical system $(\Gamma_{E},\Sigma_{E},\mu_{E},\phi_{t})$. Let $\gamma$ be a real number in $(0, 1]$ and let $1 \gg \varepsilon \ge 0$ be a very small real number so that $\gamma>\varepsilon$. If there is a macro-state $M_{V_{1}^{*}, ..., V_{l}^{*}}$ satisfying the following condition, then it is the $\gamma$-$\varepsilon$-equilibrium state of $S$: There exists a set $Y\subseteq \Gamma_{E}$ such that $\mu_{E}(Y)\geq 1-\varepsilon$ and for all initial conditions $x\in Y$:

\begin{equation}
LF_{\Gamma_{M_{V_{1}^{*}, ..., V_{l}^{*}}}}\!(x) \, \geq \,
LF_{\Gamma_{M}}\!(x)+\gamma
\end{equation}
\noindent for all macro-states $M \neq M_{V_{1}^{*}, ..., V_{l}^{*}}$. We then write $M_{\gamma\textnormal{-}\varepsilon\textnormal{-}eq}\, := \, M_{V_{1}^{*}, ..., V_{k}^{*}}$.

\end{quote}

\noindent As above, nothing in what we say about equilibrium depends on the particular value of the parameter $\gamma$ and so we leave it open what the best choice would be.\\

We contend that these two definitions provide the relevant notions of equilibrium in BSM. But the definitions remain silent about the size of equilibrium macro-regions, and they do not in any obvious way imply anything about seize. These regions being extremely small would be compatible with the definitions. That these macro-regions have the right size is a result established in the following two theorems:

\begin{quote}

\emph{Deterministic Dominance Theorem}: If  $M_{\alpha\textnormal{-}\varepsilon\textnormal{-}eq}$ is an $\alpha$-$\varepsilon$-equilibrium of system $S$, then $\mu_{E}(\Gamma_{M_{\alpha\textnormal{-}\varepsilon\textnormal{-}eq}}) \geq \beta$ for $\beta=\alpha(1-\varepsilon)$.\footnote{We assume that $\varepsilon$ is small enough so that $\alpha(1-\varepsilon)> \frac{1}{2}$.}

\end{quote}

\begin{quote}

\emph{Deterministic Prevalence Theorem}: If $M_{\gamma\textnormal{-}\varepsilon\textnormal{-}eq}$ is a $\gamma$-$\varepsilon$-equilibrium of system $S$, then $\mu_{E}(\Gamma_{M_{\gamma\textnormal{-}\varepsilon\textnormal{-}eq}}) \geq \mu_{E}(\Gamma_{M})+\gamma-\varepsilon$ for all macro-states $M\neq M_{\gamma\textnormal{-}\varepsilon\textnormal{-}eq}$.

\end{quote}

Both theorems are completely general in that no dynamical assumptions are made.\footnote{We assume that the dynamics is stationary, i.e. that $\phi_{t}$ does not depend on time explicitly. This, however, is not a substantive assumption in the current context because standard systems in statistical mechanics such as gases and crystals are stationary.} Thus the theorems also apply to strongly interacting systems. It is worth highlighting that the theorems make the conditional claim that if an equilibrium exits, then it is large in the relevant sense. Some systems have equilibria and for these the theorem holds. For instance the baker's gas (a gas consisting of $N$ copies of the baker's transformation) has an equilibrium in the requisite sense and the relevant macro-region is large (see Lavis (2005) for a discussion of the baker's gas). Other systems don't have equilibria, and for these the antecedent of the conditional is not satisfied. If, for instance, the dynamics is given by the identity function, no approach to equilibrium takes place.\\

There are many systems in statistical mechanics where a stochastic dynamics is considered.  Important examples include the Ising model, the lattice gas, the six vertex model and the eight vertex model (cf.\ Baxter 1982; Lavis and Bell 1999).  Hence the above definitions and results do not apply to them and so the question arises whether they can be carried over to stochastic systems. We now introduce stochastic systems and then show that such a generalisation is possible.

\section{Stochastic Processes}\label{Stochastic}

In order to introduce stochastic processes, we first need to define random variables. Intuitively, a random variable $Z$ gives the outcome of a probabilistic experiment, where the distribution $P\{Z\in A\}$ specifies the probability that the outcome will be in a certain range $A$. Formally, a \textit{random variable} is a measurable function $Z:\Omega\rightarrow\bar{X}$, where $(\Omega,\Sigma_{\Omega},\nu)$ is a probability space and $(\bar{X},\Sigma_{\bar{X}})$ measurable space. $\bar{X}$ is the set of possible outcomes and is therefore referred to as the \textit{outcome space}. The \textit{probability measure} $P\{Z\in A\}=\nu(Z^{-1}(A))$ for all $A\in\Sigma_{\bar{X}}$ is called the \textit{distribution} of $Z$. \\

A stochastic process consists of a string of the kind of probabilistic experiments that are described by a random variable. Formally, a \emph{stochastic process} $\{Z_{t}\}$, $t\in\field{R}$ (continuous time) or $\field{Z}$ (discrete time), is a family of random variables $Z_{t}$, which are defined on the same probability space $(\Omega,\Sigma_{\Omega},\nu)$ and take values in the same measurable space $(\bar{X},\Sigma_{\bar{X}})$ such that $Z(t,\omega)=Z_{t}(\omega)$ is jointly measurable in $(t,\omega)$. Intuitively speaking, each $\omega$ encodes an entire possible history (present and future) of a stochastic process, and thus $\Omega$ is the set of all possible histories the stochastic process (we illustrate this idea with a simple example below). A \textit{realisation} is a possible path of the stochastic process. That is, it is a function $r_{\omega}:\field{R}\rightarrow\bar{X}$, $r_{\omega}(t)=Z(t,\omega)$, for $\omega\in\Omega$ arbitrary (cf. ~Doob 1953,~4--46). The difference between $\omega$ and $r_{\omega}$  is simply that while $r_{\omega}$ gives a possible path of the stochastic process in terms of sequences of elements of $\bar{X}$, $\omega$ just \textit{encodes} such a possible history.\\

If the random variable does not depend explicitly on time (if, for instance, the outcome does not depend on when you toss a coin), then we have a stationary stochastic process and in what follows all the stochastic processes we will be working with will be assumed to be stationary. Formally: A stochastic process $\{Z_{t};\, t\in \field{Z}\}$ is \emph{stationary} iff the distributions of the multi-dimensional random variable $(Z_{t_{1}+h},\ldots, Z_{t_{n}+h})$ is the same as the one of $(Z_{t_{1}},\ldots, Z_{t_{n}})$ for all $t_{1},\ldots,t_{n}\in \field{R}$ or $\field{Z}$, $n\in\field{N}$, and all $h\in \field{Z}$ or $\field{R}$ (—\textit{ibid.}).\\

Let us now give an example, namely the discrete stochastic process that describes a bi-infinite series of coin tosses of a fair coin with probability $p_{H}=1/2$ (`Heads') and $p_{T}=1/2$ (`Tails'), $p_{H\cup T}=1$ and $p_{\emptyset}=0$. In this case $\bar{X}=\{H,T\}$ and  $\Sigma_{\bar{X}}$ is the power set of $\bar{X}$. $\Omega$ is the set encoding all possible histories of the stochastic process. That is, $\Omega$ is defined as the set of all sequences $\omega=(\ldots\omega_{-1}\omega_{0}\omega_{1}\ldots)$ with
$\omega_{i}\in\bar{X}$ corresponding to one of the possible
outcomes of the $i$-th trial in a doubly infinite sequence of
trials. $\Sigma_{\Omega}$ is the $\sigma$-algebra generated by
the cylinder-sets
\begin{equation}\label{cylinder}
C^{G_{1}...G_{n}}_{i_{1}...i_{n}}\!\!=\!\!\{\omega\in
\Omega\,|\,\omega_{i_{1}}\!\!\in\!\!G_{1},\!\ldots\!,\omega_{i_{n}}\!\!\in\!\!G_{n},
G_{j}\in\Sigma_{\bar{X}}\!,\,i_{j}\!\in\!\field{Z},\,i_{1}\!<\!\ldots\!<\!i_{n},\,1\!\leq j\!\leq\!n\}.
\end{equation}

Since the outcomes are independent, these sets have probability
$\bar{\nu}(C^{G_{1}...G_{n}}_{i_{1}...i_{n}}):=p_{G_{1}} \times \ldots \times p_{G_{n}}$. Let $\nu$ be defined as the unique extension of $\bar{\nu}$ to a measure on $\Sigma_{\Omega}$. Finally, define $Z_{t}(\omega):=\omega_{t}$ (the $t$-th coordinate of $\omega$). Then $Z_{t}(\omega)$ gives us the outcome of the coin toss at time $t$, $P\{Z_{t}=H\}=\nu(Z_{t}^{-1}(\{H\}))=1/2$ and $P\{Z_{t}=T\}=\nu(Z_{t}^{-1}(\{T\}))=1/2$ for any  $t$. Hence $\{Z_{t}\}$ is the stochastic process describing an infinite series of tosses of a fair coin, and it is also clear that this process is stationary.\footnote{Here we can also illustrate the difference between an $\omega$ and a realisation $r(\omega)$. We could, for instance, also use  `0' and `1' to encode the path of a stochastic process (where `0' encodes the outcome Heads and `1' encodes the outcome Tails). Then $\Omega$ would consist of sequences such as $\omega=(\ldots,0,1,0,1,\ldots)$, but $r(\omega)=(\ldots H,T,H,T,\ldots)$. More radically, we could also use a real number $\omega\in[0,1]$ to encode a sequence of 0s and 1s (via its binary development) and thus a sequence of outcomes of tossing a coin.}\\

\section{Equilibrium for Stochastic Processes}\label{S-Equilibrium}

Let us now return to BSM as introduced in the previous section. In the context of stochastic processes $\bar{X}$ plays the role of $\Gamma_{E}$ as giving the set of possible outcomes of the system. $Z_{t}(\omega)$ is the stochastic equivalent of $\phi_{t}(x)$ in that it gives the state of the system at time $t$. More specifically, the dynamics is determined by the probability measure $\nu$, from which transition probabilities (such as\ $P\{Z_{t}=H\,\,|$ given that $Z_{t-1}=T\}$) can be derived. These are the stochastic equivalent of $\phi_{t}(x)$ because they specify how the system evolves over time. Realisations are the stochastic equivalent of trajectories in the deterministic case in that they describe possible evolutions of the system. The probability measure $P$ defined on $\bar{X}$ is the stochastic equivalent of $\mu_{E}$ because it gives the probability of certain outcomes. Finally, the condition of stationarity is the stochastic analogue of the condition that $\mu_{E}$ is invariant in the deterministic case.\\

The macro characterisation of the system does not change, and so we consider again the macro-variables $\{v_{1}, ..., v_{k}\}$. The mathematical expression of supervenience is that the $v_{i}$ are functions on $\bar{X}$. That is, $v_{i}:\bar{X}\rightarrow \field{V}_{i}$. As above, a particular set of values $\{V_{1}, ..., V_{k}\}$ defines a \emph{macro-state} $M_{V_{1}, \ldots, V_{r}}$, and a complete set of macro-states contains all states as system can be in. Again, every macro-state $M$ is associated with a \textit{macro-region} $\bar{X}_{M}$ consisting of all $\bar{x}\in \bar{X}$ for which the system is in $M$. The definitions of prevalence and dominance carry over to the current context unchanged. That is, a macro-region $\bar{X}_{M_{eq}}$ is \emph{prevalent} iff  $P\{\bar{X}_{M_{eq}}\}>P\{\bar{X}_{M}\}+\gamma$ for some $\gamma \in (0, 1]$ for all $M\neq M_{eq}$, and $\bar{X}_{M_{eq}}$ is $\beta$-dominant
 iff $P\{\bar{X}_{M_{eq}}\}\geq \beta$  for some $\beta\in (\frac{1}{2},1]$.\\

The aim now is to carry over the above definitions of equilibrium from the deterministic to the stochastic context. To this end we first have to introduce the notion of the long-run fraction of time a realisation spends in a region $A\in \Sigma_{\bar{X}}$:

\begin{eqnarray}
LF_{A}(\omega)&=&\lim_{t\rightarrow\infty}\frac{1}{t}\int_{0}^{t}1_{A}(Z_{\tau}(\omega))d\tau\,\,\textnormal{for continuous time, i.e.}\,\,t\in\field{R},\,\,\\
LF_{A}(\omega)&=&\lim_{t\rightarrow\infty}\frac{1}{t}\sum_{\tau=0}^{t-1}1_{A}(Z_{\tau}(\omega))\,\,\textnormal{for discrete time, i.e.}\,\,t\in\field{Z}.
\end{eqnarray}

\noindent We are now in a position to state the stochastic definitions of equilibrium:

\begin{quote}

\textit{Stochastic $\alpha$-$\varepsilon$-Equilibrium.} Consider an isolated system $S$ whose macro-states are specified in terms of the macro-variables $\{v_{1}, ..., v_{k}\}$ and which, at the micro level, is a stationary stochastic process $\{Z_{t}\}$. Let $\alpha$ be a real number in $(0.5, 1]$, and let $1 \gg \varepsilon \ge 0$ be a very small real number. If there is a macro-state $M_{V_{1}^{*}, ..., V_{k}^{*}}$ satisfying the following condition, then it is the stochastic $\alpha$-$\varepsilon$-equilibrium state of $S$: There exists a set $\Omega^{*} \subseteq \Omega$ such that $\nu(\Omega^{*})\geq 1-\varepsilon$, and for all $\omega\in\Omega^{*}$:

\begin{equation}
LF_{\bar{X}_{M_{V_{1}^{*}, ..., V_{k}^{*}}}}(\omega) \geq \alpha.
\end{equation}
We then write $M_{\alpha\textnormal{-}\varepsilon\textnormal{-}eq}\, := \, M_{V_{1}^{*}, ..., V_{k}^{*}}$.
\end{quote}

\noindent The definition of the $\gamma$-$\varepsilon$-equilibrium is now straightforward:

\begin{quote}

\textit{Stochastic $\gamma$-$\varepsilon$-Equilibrium.} Consider an isolated system $S$ whose macro-states are specified in terms of the macro-variables $\{v_{1}, ..., v_{k}\}$ and which, at the micro level, is a stationary stochastic process $\{Z_{t}\}$. Let $\gamma$ be a real number in $(0, 1]$, and let $1 \gg \varepsilon \ge 0$ be a very small real number so that $\varepsilon<\gamma$. If there is a macro-state $M_{V_{1}^{*}, ..., V_{k}^{*}}$ satisfying the following condition, then it is the stochastic $\alpha$-$\varepsilon$-equilibrium state of $S$: There exists a set $\Omega^{*} \subseteq \Omega$ such that $\nu(\Omega^{*})\geq 1-\varepsilon$, and all $\omega\in\Omega^{*}$ satisfy

\begin{equation}
LF_{\bar{X}_{M_{V_{1}^{*}, ..., V_{k}^{*}}}}(\omega)\geq LF_{\bar{X}_{M}}(\omega)+\gamma
\end{equation}

\noindent for all $M\neq M_{V_{1}^{*}, ..., V_{k}^{*}}$. We then write $M_{\gamma\textnormal{-}\varepsilon\textnormal{-}eq}\, := \, M_{V_{1}^{*}, ..., V_{k}^{*}}$.

\end{quote}

The core result of this paper is that the two central theorems of the deterministic case, the Dominance Theorem and the Prevalence Theorem, have stochastic analogues. We now state the theorems and give the proof in the Appendix.

\begin{quote}

\emph{Stochastic Dominance Theorem}: If  $M_{\alpha\textnormal{-}\varepsilon\textnormal{-}eq}$ is a stochastic $\alpha$-$\varepsilon$-equilibrium of system $S$, then $P\{\bar{X}_{M_{\alpha\textnormal{-}\varepsilon\textnormal{-}eq}}\} \geq \beta$ for $\beta=\alpha(1-\varepsilon)$.\footnote{We assume that $\varepsilon$ is small enough so that $\alpha(1-\varepsilon)> \frac{1}{2}$.}

\end{quote}

\begin{quote}

\emph{Stochastic Prevalence Theorem}: If $M_{\gamma\textnormal{-}\varepsilon\textnormal{-}eq}$ is a stochastic $\gamma$-$\varepsilon$-equilibrium of system $S$, then $P\{\bar{X}_{M_{\varepsilon\textnormal{-}eq}}\} \geq P\{\bar{X}_{M}\}+\gamma-\varepsilon$ for all macro-states $M$ with $M\neq M_{\gamma\textnormal{-}\varepsilon\textnormal{-}eq}$.

\end{quote}

As in the deterministic case, both theorems are completely general in that no dynamical assumptions are made and hence the theorems apply to stochastic process with any dynamics.\footnote{We assume that the dynamics is stationary, but, as in the deterministic case, this is not a substantive assumption because standard stochastic systems in statistical mechanics are stationary.} As in the deterministic case it is worth noting that the theorems make the conditional claim that if an equilibrium exits, then it is large in the relevant sense. There are processes that do not have an equilibrium. For instance, consider the stochastic process of throwing a fair die (with six sides). Suppose that the macro-variable of concern is whether the die shows an even number (2, 4, 6) or an odd number (1, 3, 5). Then there will be no equilibrium because for almost any realisation half of the time the dice will show an even number and half of the time they will show an odd number.\\

\section{Example: The Lattice Gas}\label{Example}

We now illustrate the definitions and theorems of the previous section with the lattice gas. The lattice gas is a popular model not only of gases (as its name would suggest), but in fact also of liquids and solids.\footnote{See Baxter (1982) and Cipra (1987) for more details about the lattice gas.} The lattice gas models a fluid in the sense that flows are represented by particles moving from site to site, and because the system is in contact with an energy and particle reservoir, particles can also be created and annihilated. More specifically, consider a lattice with $N\in\field{N}$ sites. Each lattice site can either be occupied by a particle or be empty. This is formalised by associating with every lattice site $i$ a variable $s_{i}$, which takes the value 1 if the site is occupied and 0 if the site is empty. Thus the micro-state of the lattice is a vector $s=(s_1,\ldots,s_N)$, specifying  which sites are occupied and which ones are empty. Hence the system's $\bar{X}$ consists of the $2^N$ possible arrangements of different numbers of particles on the $N$ sites, and  $\Sigma_{\bar{X}}$ is the power set of $X$. Now the elements of $\Omega$ encode the history, present and future of the stochastic process in all its details (for the lattice gas discrete time steps are considered). That is, $\Omega$ consists of all bi-infinite sequences $\omega=(\ldots\omega_{-1}\omega_{0}\omega_{1}\ldots)$ where the $i$-th coordinate $\omega_{i}$ is an arbitrary vector $s$. $\Sigma_{\Omega}$ is the $\sigma$-algebra generated by cylinder sets that are described in Equation \ref{cylinder} if we replace the $G$s by $B$s. Finally, $Z_{t}(\omega):=\omega_{t}$ (the $t$-th coordinate of $\omega$).\\

The probability measure $\nu$ depends on the exact stochastic dynamics of the system. Many different kinds of stationary stochastic dynamics are considered for the lattice gas model (cf. Baxter 1982; Cipra 1988). At this point it is not necessary to commit to any specific stochastic dynamics. It suffices to say that a stochastic dynamics will determine the measure assigned to the cylinder sets $\bar{\nu}(C^{B_{1}...B_{n}}_{i_{1}...i_{n}})$ and its unique extension $\nu$. What we need to mention, however, is that the potential energy and the grand-canonical probability distribution will constrain the dynamics. The simplest still somewhat realistic potential is the so-called square-well potential, where only nearest neighbour interactions are taken into account. The underlying idea is that there cannot be two particles on the same site, that particles are attracted when they are close to each other and that no interaction takes place when they are far apart.

\begin{eqnarray}\label{NNI}
\phi(i,j) & = & \left\{
\begin{array}{ll}
\infty & \mbox{if $i=j$} \\
-\xi & \mbox{if $i,j$ are nearest neighbours}\\
0 & \mbox{otherwise}
\end{array}
\right\},
\end{eqnarray}

\noindent where $i$ and $j$ denote sites of the lattice and $\xi>0$. The total potential energy of the system is given by $E(s)=\sum_{i,j}\phi(i,j)s_{i}s_j$, where the sum is over all pairs of sites on the lattice (with $i\neq j$).\\

The probability measure of a set of micro-states $A$ is given by the grand-canonical probability distribution $P\{A\}$. This distribution depends on the effective chemical potential $\mu_{c}$ (one can think of the chemical potential as a measure for how accepting the system is of new particles, or for how much work one has to do to add a new particle to the system):

\begin{equation}\label{probLG}
\frac{\sum_{s\in A}\exp[\frac{\xi}{4kT}(\sum_{\textnormal{all}\,\,i,j}(2s_{i}-1)(2s_{j}-1)+
(2\mu_{c}+q\xi)\sum_{all\,\,i}(2s_{i}-1)+N(\frac{1}{2}q\xi+2\mu_{c}))]}{
\sum_{\textnormal{all}\,\,s}\exp[\frac{\xi}{4kT}(\sum_{\textnormal{all}\,\, i,j}(2s_{i}-1)(2s_{j}-1)+(2\mu_{c}+q\xi)\sum_{all\,\, i}(2s_{i}-1)+N(\frac{1}{2}q\xi+2\mu_{c}))]},
\end{equation}
where $k$ is the Boltzmann constant, $T$ is the temperature and $q$ is the number of nearest-neighbours. For any stochastic dynamics that satisfies the constraints that the potential energy is given by Equation (\ref{NNI}) and that the probabilities are specified by the grand-canonical partition function, $\{Z_{t}\}$ is a stochastic process describing the lattice gas.\\

The \emph{macro-states} usually considered are defined by the average particle density per site over the entire system: $M^{LG}_j=j/N$ where $j$ denotes the total number of molecules $s_{1}+s_2\ldots+s_N$. The \emph{macro-regions} $\bar{X}_{M^{LG}_j}$ are defined as the set of micro-states for which the system is in  macro-state $M^{LG}_j$. \\

The behaviour of the lattice gas depends on the values of the various parameters. For the purpose of illustrating our ideas, we will consider two kinds of behaviour (corresponding to ranges of parameter values). First, consider a sufficiently large $\mu_{c}$ (that is, when the system readily accepts new particles). In this case, under the usual stochastic dynamics considered, the system will spend most of the time in the macro-state where all sites are occupied, i.e.\ in $M^{LG}_N$, for almost all initial states (in a measure-theoretic sense) (cf. \ Baxter 1982). For this reason $M^{LG}_N$ is a $\gamma$-$0$ equilibrium. Thus, by the Stochastic Prevalence Theorem,  $M^{LG}_N$ is $\gamma$-prevalent.\footnote{Note that it is also clear from Equation~(\ref{probLG}) that for sufficiently large $\mu_{c}$, $M^{LG}_N$ corresponds to the largest macro-region.}\\

Second consider a sufficiently small negative-valued $\mu_{c}$ (in which case the system tends to annihilate particles). Then, under the usual stochastic dynamics considered, the system will spend most of the time in the macro-state where all sites are empty, i.e.\ in $M^{LG}_0$, or almost all initial states (cf.\ Baxter 1982). Therefore, $M^{LG}_0$ is a $\gamma$-$0$-equilibrium. Thus, by the Stochastic Dominance Theorem,  $M^{LG}_N$ is $\gamma$-prevalent.\footnote{Again, this is clear from Equation~(\ref{probLG}).}\\

To conclude, \emph{the lattice gas represents an important physical system that has equilibria in our sense}. Let us end with a few remarks on why this system is physically important. First, it provides a good model of \emph{condensation} and the \emph{liquid-gas transition}.\footnote{Mathematically speaking, the lattice gas is equivalent to the \emph{Ising model}. The Ising model is one of the best developed and most widely studied models in physics and is discussed in nearly every modern textbook on statistical mechanics. In particular, the lattice gas on a square lattice with $\mu_{C}=-\xi/8$ is equivalent to the two-dimensional Ising model with no external field, which is famous for being one of the very few exactly solved models that display \emph{phase transitions} (Baxter 1982).} A theory of condensation was developed based on the lattice model, which was shown to qualitatively reproduce the main features of condensation and was empirically confirmed for many cases (Kierlik et al.\ 2002; Young and Lee 1952). To give an example, De Ribaupierre and Manchester (1974) found that the lattice gas provides a satisfactory model of condensation for a hydrogen in palladium system. Pan et al.\ (1998) found that the lattice gas gives a fair description of the liquid-gas transition in excited nuclear systems formed as a result of a heavy ion collisions. Finally, the lattice gas also models \emph{melting and freezing} phenomena well (see Kikuchi and Cahn 1980). For instance, Clarke et al.\ (1979) found that the lattice gas model provides a good description of melting for graphite intercalated with caesium.\\

\section{Conclusion}\label{Conclusion}
We presented stochastic formulations of the notions of an $\alpha$-$\varepsilon$-equilibrium and a $\gamma$-$\varepsilon$-equilibrium, and we have formulated and proven stochastic equivalents of the Dominance Theorem and the Prevalence Theorem. This completes the transfer of the basic notions of our framework from the deterministic to the stochastic context. There is, however, an important disanalogy between the two contexts as far as the existence of an equilibrium state is concerned. In the deterministic context we were able to prove an existence theorem (Werndl and Frigg 2015, pp. 26-29). There is no straightforward generalisation of this theorem to the stochastic context. This is because the conditions that need to hold for an equilibrium to exist in the existence theorem are conditions on the ergodic components. However, stochastic processes do not have such ergodic components. It is true that the deterministic representation of a stochastic process (cf. the Appendix for a definition) has ergodic components. However, these are components of $\Omega$ and not of $\bar{X}$, and an existence theorem would need to be about $\bar{X}$. Thus, as far as we see, the ergodic components of $\Omega$ are not useful to characterise the circumstances under which equilibria exist for stochastic processes. Hence there is an open question about when a stochastic equilibrium exists.

\pagebreak


\section{Appendix}

\subsection{Proof of the Stochastic Dominance Theorem}\label{Appendix1}

First of all, let us show that a stationary stochastic process $\{Z_{t}\}$ can be represented by a measure-preserving deterministic system $(X,\Sigma_{X},\mu_{X},T_{t})$. Let $X$ be the set of all possible realisations, i.e., functions $x(\tau)$ from $\field{R}$ or $\field{Z}$ to $\bar{X}$. Let $\Sigma_{X}$ be the $\sigma$-algebra generated by the cylinder-sets
\begin{equation}\label{cylinder}
C^{A_{1}...A_{n}}_{i_{1}...i_{n}}\!\!=\!\!\{x\!\in\! X\,|\,x(i_{1})\!\!\in\!\! A_{1},...,x(i_{n})\!\!\in\!\! A_{n}, A_{j}\!\in\! \Sigma_{\bar{X}},i_{j}\!\in\!\field{R}\,\,\textnormal{or}\,\, \field{Z},\,i_{1}\!\!<...<\!\!i_{n},1\!\leq j\!\leq n\}.
\end{equation}

Let $\mu_{X}$ be the unique probability measure arising by assigning to each  $C^{A_{1}...A_{n}}_{i_{1}...i_{n}}$ the probability $P\{Z_{i_{1}}\in A_{1},\ldots,Z_{i_{n}}\in A_{n}\}$. The evolution functions shift a realisation $t$ times to the left, i.e., $T_{t}(x(\tau))=x(\tau+t)$. The $T_{t}$ are invariant under the dynamics because $\{Z_{t}\}$ is stationary. $(X,\Sigma_{X},\mu_{X},T_{t})$ is a measure-preserving deterministic system called the \emph{deterministic representation} (cf.~Doob 1953,~621--622; Werndl 2009, 2011).\\

Let $W=\{x(\tau)\in X\,\,|\,\,x(\tau)=Z_{\tau}(\omega)$ for all $\tau$ for a $\omega\in\Omega^{*}\}$. Note that $\mu_{X}(W)\geq1-\varepsilon$. Consider first the case of an $\alpha$-$\varepsilon$-equilibrium $M_{\alpha\textnormal{-}\varepsilon\textnormal{-eq}}$. Then it follows that for all $x\in W$:
\begin{equation}\label{wuzi}
LF_{X_{Q_{M_{\alpha\textnormal{-}\varepsilon\textnormal{-eq}}}}}(x)\geq \alpha,
\end{equation}
where $Q_{M_{\alpha\textnormal{-}\varepsilon\textnormal{-eq}}}=\{x\in X\,\,|\,\,x(0)\in \bar{X}_{M_{\alpha\textnormal{-}\varepsilon\textnormal{-eq}}}\}$.\\

Hence $Q_{M_{\alpha\textnormal{-}\varepsilon\textnormal{-eq}}}$ is an $\alpha$-$\varepsilon$-equilibrium of $(X,\Sigma_{X},\mu_{X},T_{t})$. It follows from the (deterministic) Dominance Theorem (Frigg and Werndl 2015a) that $\mu_{X}(Q_{M_{\alpha\textnormal{-}\varepsilon\textnormal{-eq}}})> \alpha(1-\varepsilon)$, which immediately implies that $P\{M_{\alpha\textnormal{-}\varepsilon\textnormal{-eq}}\}> \alpha(1-\varepsilon)$.\\

\subsection{Proof of the Stochastic Prevalence Theorem}\label{Appendix}

The proof proceeds in the same fashion as the previous one. That is, consider again the measure-preserving deterministic system $(X,\Sigma_{X},\mu_{X},T_{t})$ that represents the stationary stochastic process $\{Z_{t}\}$. Suppose that $M_{\gamma\textnormal{-}\varepsilon\textnormal{-eq}}$ is an $\gamma$-$\varepsilon$-equilibrium. \\

As before, let $W=\{x(\tau)\in X\,\,|\,\,x(\tau)=Z_{\tau}(\omega)$ for all $\tau$ for a $\omega\in\Omega^{*}\}$. Note that $\mu_{X}(W)\geq 1-\varepsilon$.\\

Then for all $x\in W$ and all $M\neq M_{\gamma\textnormal{-}\varepsilon\textnormal{-eq}}$ it holds that
\begin{equation}LF_{X_{Q_{M_{\gamma\textnormal{-}\varepsilon\textnormal{-eq}}}}}(x)\geq LF_{X_{Q_{M}}}+\gamma-\varepsilon,
\end{equation}
where
$Q_{M_{\gamma\textnormal{-}\varepsilon\textnormal{-eq}}}=\{x\in X\,\,|\,\,x(0)\in \bar{X}_{M_{\gamma\textnormal{-}\varepsilon\textnormal{-eq}}}\}$ and
$Q_{M}=\{x\in X\,\,|\,\,x(0)\in \bar{X}_M\}$.
Hence $Q_{M_{\gamma\textnormal{-}\varepsilon\textnormal{-eq}}}$ is an $\gamma$-$\varepsilon$-equilibrium of $(X,\Sigma_{X},\mu_{X},T_{t})$. \\

It follows from the (deterministic) Prevalence Theorem (cf.\ Werndl and Frigg 2015a) that
$\mu_{X}(Q_{M_{\gamma\textnormal{-}\varepsilon\textnormal{-eq}}})\geq\mu_{X}(Q_{M})+\gamma-\varepsilon$ for all $M\neq M_{\gamma\textnormal{-}\varepsilon\textnormal{-eq}}$. This immediately implies that $P\{M_{\gamma\textnormal{-}\varepsilon\textnormal{-eq}}\}\geq
P\{M\}+\gamma-\varepsilon$ for all $M\neq M_{\gamma\textnormal{-}\varepsilon\textnormal{-eq}}$.

\section*{References}

\noindent Baxter, Rodney. 1982. \emph{Exactly Solved Models in Statistical Mechanics}. San Diego: Academic Press Limited.\\

\noindent Callender, Craig. 2001. Taking Thermodynamics Too Seriously. \emph{Studies in History and Philosophy of Modern Physics} 32: 539-553.\\

\noindent Cipra, Barry A. (1987). An Introduction to the Ising Model. \textit{American Mathematical Monthly} 94: 937-954.\\

\noindent Clarke, R., N. Caswell, and S. A. Solin. 1979. Melting and Staging in Graphite Intercalated with Cesium. \emph{Physical Review Letters} 42: 61-64. doi:http://dx.doi.org/\\\-10.1103/PhysRevLett.42.61.\\

\noindent de Ribaupierre, Y., and F. D. Manchester. 1974. Experimental study of the critical-point behaviour of the hydrogen in palladium system. I. Lattice gas aspects. \emph{Journal of Physics C: Solid State Physics}  7: 2126-2139. doi:10.1088/0022-3719/7/12/006.\\

\noindent Doob, Joseph L. 1953. \emph{Stochastic Processes}. New York: John Wiley \& Sons.\\

\noindent Frigg, Roman. 2008. A Field Guide to Recent Work on the Foundations of Statistical Mechanics. In \emph{The Ashgate Companion to Contemporary Philosophy of Physics}, ed. Dean Rickles, 99-196. London: Ashgate.\\

\noindent Kierlik, E., P. A. Monson, M. L. Rosinberg, and G. Tarjus. 2002. Adsorption hysteresis and capillary condensation in disordered porous solids: a density functional study. \emph{Journal of Physics: Condensed Matter} 14: 9295-9315. doi:10.1088/0953-8984/14/40/319.\\

\noindent Kikuchi, Ryoichi,  and John W. Cahn. 1980. Grain-boundary melting transition in a two-dimensional lattice-gas model. \emph{Physical Review B} 21: 1893-1897.\\

\noindent Lavis, David. 2005. Boltzmann and Gibbs: An attempted reconciliation. {\it Studies in History and Philosophy of Modern Physics} 36: 245-273.\\

\noindent Lavis, David, and George M. Bell. 1999. \emph{Statistical Mechanics of Lattice Systems: Volume 1: Closed-Form and Exact Solutions}. Berlin: Springer.\\

\noindent Jicai, Pan, Subal Das Gupta, and Martin Grant. 1998. First-Order Phase Transition in Intermediate-Energy Heavy Ion Collisions. \emph{Physical Review Letter} 80: 1182-1885.\\

\noindent Reiss, Howard. 1996. \emph{Methods of Thermodynamics}. Mineaola/NY: Dover.\\

\noindent Wang, Genmiao, Edith M. Sevinck,  Emil Mittag, Debra J. Searles,  and Denis J.\ Evans. 2002. Experimental Demonstration of Violations of the Second Law of Thermodynamics for Small Systems and Short Time Scales. \emph{Physical Review Letters} 89: 050601.\\

\noindent Werndl, Charlotte. 2009. Are Deterministic Descriptions and Indeterministic Descriptions Observationally Equivalent?. \emph{Studies in History and Philosophy of Modern Physics} 40: 232-242.\\

\noindent Werndl, Charlotte. 2011. On the Observational Equivalence of Continuous-Time Deterministic and Indeterministic Descriptions. \emph{European Journal for the Philosophy of Science} 1: 193-225.\\

\noindent Werndl, Charlotte, and  Roman Frigg. 2015a. Reconceptualising Equilibrium in Boltzmannian Statistical Mechanics and Characterising its Existence. \emph{Studies in History and Philosophy of Modern Physics} 44 : 470-479.\\

\noindent Werndl, Charlotte, and Roman Frigg. 2015b. Rethinking Boltzmannian Equilibrium. \emph{Philosophy of Science} 82: 1224-1235.\\

\noindent Yang, Chen Ning, and Tsung-Dao Lee. 1952. Statistical Theory of Equations of State and Phase Transitions. I. Theory of Condensation. \emph{Physical Review} 87: 404-409.

\end{document}